\documentclass[amsmath,amssymb,aps,prl,twocolumn,groupedaddress,superscriptaddress,10pt,reprint,longbibliography]{revtex4-2}

\usepackage[left]{lineno}
\usepackage{graphicx}
\usepackage{textcomp} 
\usepackage{bm}

\usepackage{exscale}

\usepackage{relsize}

\usepackage{graphicx}
\usepackage{dcolumn}
\usepackage{bm}
\usepackage[colorlinks, linkcolor=blue, citecolor=blue, urlcolor=blue, breaklinks=true]{hyperref}
\usepackage{color}
\usepackage{float}
\usepackage[
margin=0.71in,
]{geometry}

\begin{document}

\title{Observation of Janus Chirality for Coherent Thermal Emission from Metasurfaces}

\author{Kaili Sun}
\affiliation{Shandong Provincial Key Laboratory of Light Manipulation and Applications, School of Physics and Opto-Electronics, Shandong Normal University, Jinan 250358, China}

\author{Yangjian Cai}
\affiliation{Shandong Provincial Key Laboratory of Light Manipulation and Applications, School of Physics and Opto-Electronics, Shandong Normal University, Jinan 250358, China}

\author{Maxim V. Gorkunov}
\email{gorkunov@crys.ras.ru}
\affiliation{National University of Science and Technology MISIS, Moscow 119049, Russia}
\affiliation{
Shubnikov Institute of Crystallography, 
Moscow 119333, Russia}

\author{Yuri Kivshar}
\email{yuri.kivshar@anu.edu.au}
\affiliation{Research School of Physics, Australian National University, Canberra ACT 2615, Australia}
\affiliation{Department of Physics, The University of Hong Kong, 999077 Hong Kong, China }

\author{Zhanghua Han}
\email{zhan@sdnu.edu.cn}
\affiliation{Shandong Provincial Key Laboratory of Light Manipulation and Applications, School of Physics and Opto-Electronics, Shandong Normal University, Jinan 250358, China}


\begin{abstract}
Metasurfaces emerged as a powerful tool for controlling thermal radiation, yet achieving coherent emission with opposite circular handednesses remains a highly challenging problem. Here, we demonstrate experimentally the Janus chiral thermal emission from metasurfaces with opposite circular handednesses on either side of a single device. We employ anisotropic metasurfaces supporting high-$Q$ resonances with photonic flatbands enabling near-unity circular dichroism through in-plane symmetry control. Our experiments confirm the Janus coherent emission, and they are validated by the results of the coupled-mode theory. The flatband resonant metasurfaces enabling a control of chiral thermal emission provide an efficient platform for spin-controlled light-matter interaction. 
\end{abstract}

\maketitle

\textit{Introduction}—With a rapid development of metasurface technology, precise control of the polarization state of light, the direction of propagation, and the local density of states through artificial structures has become a key research frontier \cite{Huang2023Resonant, Brongersma2025The,Gorkunov2020Metasurfaces,Shi2022Planar,Wang2024Optical}. In recent years, nonreciprocal and Janus metasurfaces have emerged as innovative approaches for precise control of different radiation channels \cite{Cotrufo2023nonreciprocal,Zhu2021Janus,Wang2025Terahertz,Chen2019Directional, Chen2023Directional,Picardi2018Janus}. Previous studies have shown that the far-field polarization vector fields of radiations from photonic crystal slabs and metasurfaces display rich topological features, such as bound states in the continuum (BICs) \cite{Kang2025Janus,Yin2025Janus,Zuo2025Janus,Song2025Parity,Ma2026Janus}, unidirectional guided resonances (UGRs) \cite{Ji2026Janus}, and circularly polarized states (C-points) \cite{Lin2025Chiral,Wang2025High,Zhao2024Spin,Lv2024Robust}. However, existing studies often rely on complex geometries with both in-plane and out-of-plane mirror symmetry-breaking or introduce asymmetry through magneto-optic materials \cite{Lv2024Robust} and anisotropic crystals \cite{Ji2026Janus}. These methods not only necessitate complicated fabrication processes with high precision requirements but are also challenging to control experimentally, resulting in relevant experimental reports being scarce. Thus recent studies have shown strong resonant optical chirality can be achieved in structures fabricated by tilted etching techniques \cite{Chen2023Observation,Zhang2022Chiral} and by multi-step variable-height lithography \cite{Kuehner2023, Heimig2024}.
In flat structures which are more convenient for fabrication, circularly polarized radiation is usually confined to off-$\Gamma$-point regions \cite{Liu2019Circularly}. 
In the most cases, circularly polarized eigenstates provide 
homochiral output of the same handedness in the upward and downward directions.

For thermal emission, traditional blackbody radiation exhibits typically disordered and isotropic characteristics. However, by manipulating the local density of states of photons through artificial nanostructures, precise control over the emission spectrum, direction, and polarization can reshape thermal fluctuations to become temporally and spatially coherent~\cite{Overvig2021Thermal, Greffet2002Coherent, DeZoysa2012Conversion, Nolen2024Local, Sun2025Full, Wang2023Observation, Inoue2014Realization}. Nevertheless, conventional devices typically enable the emission of light of a single helicity at a specific frequency and in a narrow range of directions, which requires integration of multiple devices for practical applications, thus significantly increasing the footprint. These limitations hinder the scalability and compactness necessary for advanced technologies such as infrared imaging \cite{Chen2025Thermal}, optical communication \cite{Miyoshi2018High}, and sensing \cite{Tan2020Non}.

In this Letter, we introduce the concept of Janus chiral thermal emission where a single device demonstrates coherent wide-angle circularly polarized thermal emission of opposite handedness in the upper and lower half-spaces. We demonstrate this concept for anisotropic metasurfaces, composed of arrays of weakly coupled waveguides with lateral periodic modulation on both sides. This platform is related to the coupled-resonator optical waveguides (CROWs)~\cite{Sun2025Flatband,Yariv1999} enabling the simultaneous control of a pair of thermal photons carrying opposite spin angular momenta. The structure supports tunable high-$Q$ resonances, and demonstrates wide-angle photonic flatband behavior. Notably, the chiral Janus emission is immune to variations in the substrate’s refractive index and persists under both intact and broken $\sigma_z$ symmetry conditions. Experimental results validate these behaviors, showcasing thermal emission with near-unity CD ($ECD > 0.8$) and large temporal coherence ($Q > 150$), while featuring opposite handedness confirmed on different sides of the fabricated sample. This approach streamlines the fabrication process while providing a versatile route for multifunctional chiral light sources and sophisticated radiative control.

\begin{figure*}
\centering
\includegraphics[width=0.7\textwidth]{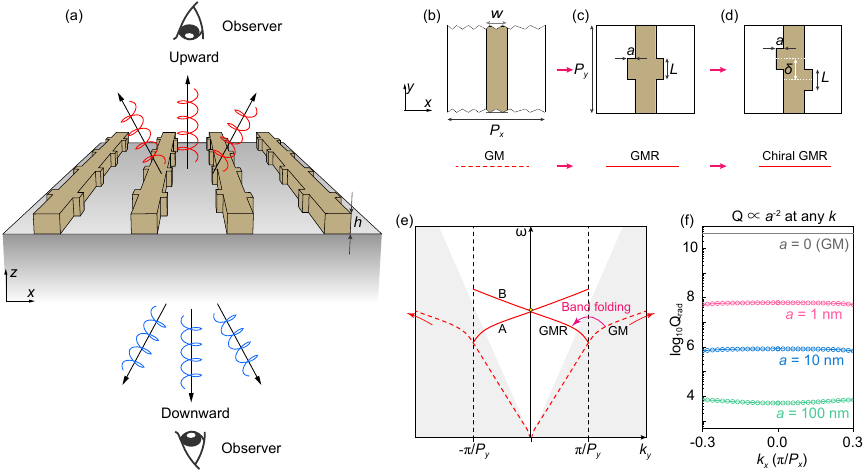}
\caption{(a) Schematic of the principle of Janus chirality from CWMs, where red and blue beams represent upward- and downward-directed circularly polarized emission of opposite handedness, respectively. (b–d) Evolution of the structural design from a conventional 1D waveguide array to a chiral CWM. (e) Band folding in the $k_y$ direction resulting from the transition from a 1D lattice of uncoupled waveguides to a periodic structure with period $P_y$. (f) $Q_{rad}$-factors of band A versus $k_y$ for various corrugation depths $a$ (0, 1, 10, and 100 $nm$) as depicted in (d).}
\label{fig1}
\end{figure*}

The operational principle of the Janus chirality in CWMs is schematically illustrated in Fig.\ref{fig1}(a), where the red and blue beams represent thermal emissions with opposite handedness propagating into upward and downward directions, respectively. For our design, we employ a low-loss CaF$_2$ substrate ($n$ = 1.4) and a CWM made of evaporated Ge (refractive index \( n = 4.2048+0.002i \) at \( T = 200^\circ \text{C} \)). This material platform is specifically suitable for thermal emission in the mid-infrared (MIR), a spectral range of paramount importance due to its rich molecular fingerprinting capabilities. 

We begin with a periodic optical waveguide array that provides strong modal confinement, as suggested by a series of guided modes (GMs) entirely below the light cone [red dashed line, Fig.\ref{fig1}(e)]. By incorporating identical lateral sawtooth corrugations of the depth $a$ on both sides of the waveguides [see Fig.\ref{fig1}(c)], the translational symmetry along the $y$-direction is broken. This perturbation enables GMs originally confined within each waveguide to interact with the radiative continuum, giving rise to a series of high-$Q$ quasi-BICs. These states, which extend spatially along the $y$-direction, exhibit nonlocal characteristics and are manifested as guided-mode resonances (GMR), as indicated by the solid red line in Fig.\ref{fig1}(e).

When a central distance $\delta$ is introduced between adjacent sawtooth corrugations on opposite sides [see Fig.\ref{fig1}(d)], the in-plane mirror symmetry of the structure is broken, while the $C_2$ symmetry is preserved. This symmetry breaking enables the realization of chiral GMR. For the target emission wavelength in the MIR, we consider the following geometric parameters: $P_x = 3000$~nm, $P_y = 2380$~nm, $h = 515$~nm, $L = 535$~nm, $w = 980$~nm, and $\delta = 735$~nm.

Fig.\ref{fig1}(f) presents the radiative Q-factor $Q_{\rm rad}$ of band A as a function of $k_x$ calculated for lossless CWMs with different corrugation depth $a$. The results indicate that, for a given $a$, $Q_{\rm rad}$ remains stable across the momentum space. Importantly, for any specific $k$ value, it follows an inversely quadratic relationship with $a$, i.e., $Q_{\rm rad} \propto 1/a^2$ \cite{Sun2023Infinite,Sun2024High}. This behavior contrasts sharply with traditional BICs, where the Q-factor is highly $k$-dependent and typically exhibits a rapid decay, following a relation $Q \propto 1/k^{\beta}$ with a constant $\beta$ \cite{Jin2019Topologically,Kang2021Merging}, which results in high-Q resonances confined to a very narrow range of angles of incidence and emission.

\begin{figure*}
\centering
\includegraphics[width=0.7\textwidth]{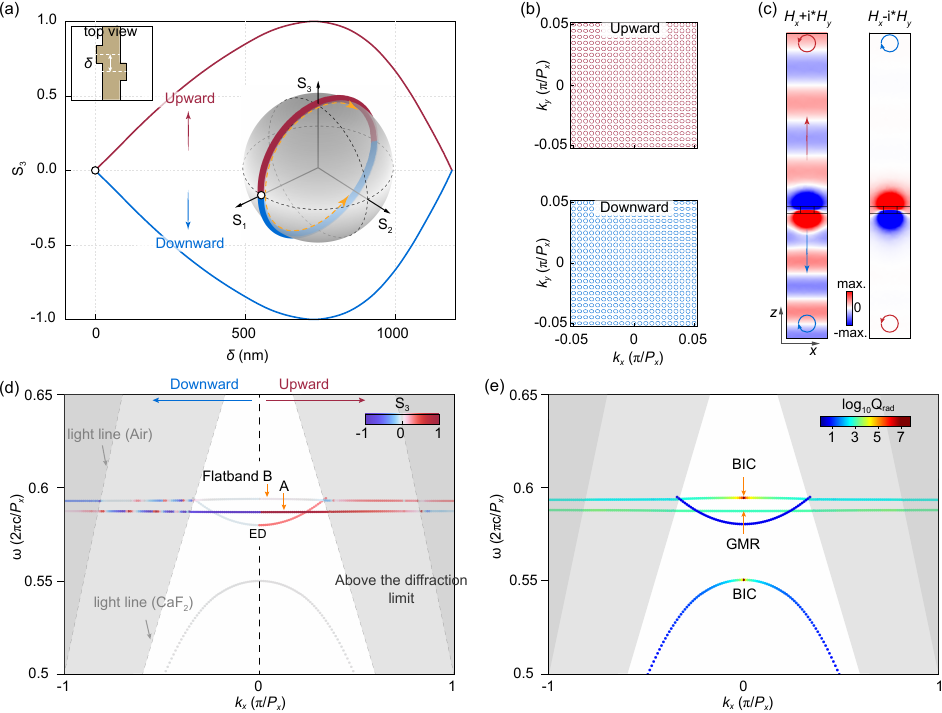}
\caption{(a) Evolution of Stokes parameter \(S_3\) for upward and downward emission as a function of $\delta$. At $\delta$ = 735 $nm$, \(S_3 = \pm 1\). Left inset: top view of the structure; right inset: the trajectory of polarization states on the Poincaré sphere as \(\delta\) changes. (b) Simulated far-field polarization maps in momentum space for upward and downward radiations at $\delta$ = 735 $nm$. (c) Circular polarization components in the \(xz\) plane at the \(\Gamma\) point. (d) Calculated band structure at $\delta$ = 735 $nm$, with the color scale indicating \(S_3\) values for upward and downward radiations. (e) Q factor of the band structure at $\delta$ = 735 $nm$.}
\label{fig2}
\end{figure*}

According to the temporal coupled-mode theory (TCMT), the absorptance $A$ and emissivity $E$ of a resonant dual-port system can be written as \cite{Haus1984,Sun2024Ultra}:
\begin{equation}
E(\omega) = A(\omega) = \frac{2\gamma \gamma_d}{(\omega - \omega_0)^2 + (\gamma + \gamma_d)^2},
\end{equation}
where \(\omega\) is the frequency of the thermal emission, \(\omega_0\) is the resonance frequency of the structure, \(\gamma_d\) is the resonance dissipative loss rate, and \(\gamma\) is the radiative decay rate. When the critical coupling condition \(\gamma = \gamma_d\) is met, at the resonance \(\omega = \omega_0\), the emissivity can be strictly enhanced to 0.5. While the absorption Q-factor \(Q_{\text{abs}} = {\omega_0}/{2\gamma_d}\) is determined by the intrinsic loss in the CWM material, the radiative Q-factor \(Q_{\text{rad}} = {\omega_0}/{2\gamma}\) is controlled by the corrugation depth $a$. We empirically match \(Q_{\text{rad}}=Q_{\text{abs}}\) by selecting $a = 210$~nm.

To better understand the impact of \(\delta\) on chirality, we calculate the variation of the Stokes parameters for the upward and downward far-field radiation of band A as a function of \(\delta\), with the results summarized in Fig.~\ref{fig2}(a). The Stokes parameters used to characterize the polarization state are defined as \cite{Ye2020Singular,Liu2019Circularly}: $S_0 = |c_x|^2 + |c_y|^2$, $S_1 = |c_x|^2 - |c_y|^2$, $S_2 = 2\Re(c_x c_y^*)$, $S_3 = i(c_x c_y^* - c_y c_x^*)$, where $c_x$ and $c_y$ are the complex-valued far-field polarization vectors extracted from the eigenmode calculations \cite{Zhen2014Topological}. As \(\delta\) increases from zero, \(|S_3|\) grows keeping opposite signs for the upward and downward directions. At $\delta = 735$~nm, it reaches the extrema with \(|S_3|=1\), and, as $\delta$ further increases, gradually decreases until vanishing at $\delta$ = $P_y/2 = 1190$~nm when the structure regains its in-plane mirror symmetry. The inset shows the trajectory of motion of the far-field Stokes parameters for both directions upon the Poincar\'e sphere, the possibility of full ellipticity coverage.

The far-field polarization maps for both directions at $\delta = 735$~nm are shown in Fig.\ref{fig2}(b), where the red and blue colors represent the left- and right-handed circular polarization (LCP and RCP) output, respectively. Our findings reveal that the Janus chiral emission is not only characterized by a strict side-handedness correlation but is also remarkably robust across a substantial portion of the Brillouin zone. Rather than being restricted to an isolated singularity--as it is typical for chiral BICs--the C-points in our platform amount to a continuous C-area in the momentum space (see Section 5 of the Supplemental Information). This topological feature facilitates near-unity CD over an extensive angular range, underscoring the versatility of the CWM architecture in achieving sophisticated, wide-angle control of the emission chirality.

Fig.\ref{fig2}(c) illustrates the computed eigenmode profile at the \(\Gamma\)-point with $\delta = 735$~nm, focusing on the \(H_x \pm iH_y\) field distributions in the $xz$-plane. Our results show that the emission in both directions is dominated by the \(H_x + iH_y\) component, representing LCP and RCP for upward and downward radiations, respectively. This handedness flip is consistent with the standard convention, where the sign of circular polarization is defined relative to the propagation direction. The near-zero intensity of the \(H_x - iH_y\) component for both radiative directions unequivocally validates the Janus chiral nature of the emission. 

We further calculated the band structures of band A and B along the \(k_x\) direction, as well as the \(S_3\) values for the upward and downward radiations using the calculated eigenmode distributions, based on the aforementioned geometric parameters, as shown in Fig.\ref{fig2}(d). The opposite \(S_3\) values, with absolute values close to 1, span a wide wavevector range in band A, consistent with the results shown in Fig.\ref{fig2}(b), thus confirming the characteristic features of C-plane. Notably, these modes exhibit photonic flatband behavior along the \(k_x\) direction. This phenomenon primarily arises from the CROW effect, a concept originally introduced by Yariv et al. \cite{Yariv1999} in microcavity arrays and later applied by Sun et al. \cite{Sun2025Flatband} in nonlocal metasurfaces. In this design, each corrugated waveguide can be viewed as an individual high-$Q$ resonator(see Fig.S2). Unlike traditional discrete coupled resonators, the quasi-guided Bloch modes in our system result from the periodic modulation of the waveguide structure, which forms a unique set of GMR states that can be manipulated in a similar yet distinct manner.

Assuming the coupling strength between the neighboring resonators is \(\kappa\), in the tight-binding approximation the dispersion relation can be written as:
\begin{equation}
\omega(k) = \omega_0 + 2\kappa \cos(kP_x),
\end{equation}
where \(\omega_0\) is the high-$Q$ resonance frequency of each GMR. The \(\cos(kP_x)\) term reflects the periodic coupling between resonators and the interference of the light modes when they propagate through the periodic array. This equation clearly shows that for smaller \(\kappa\) and larger \(P_x\), a flatter dispersion relation forms. The coupling strength \(\kappa\) typically relates to the period by \(\kappa \propto 1/P_x^\alpha\), where \(\alpha\) is a constant, implying that the coupling strength decreases as the period increases. We also calculated the corresponding \(Q_{\text{rad}}\) factors for both bands under this geometric configuration, as shown in Fig.\ref{fig2}(e). For band A, the overall \(Q_{\text{rad}}\) remains stable above \( 10^3 \). In contrast, for band B, the formation of the \(\Gamma\)-point BIC due to the rotational symmetry of the modes is observed, with additional results in Fig.S3 showing that this aligns with the traditional Janus BIC characteristics. However, for the lower-frequency ($\sim0.55 \times 2\pi c/ P_x$) dispersion BIC shown in Figs.\ref{fig2}(d) and (e), it exhibits the traditional same radiation characteristics for both upward and downward directions (see Fig.S4).

Crucially, our findings challenge the conventional paradigm, which generally mandates the breaking of \(\sigma_z\) mirror symmetry to achieve chiral optical response. In our case, within the range of substrate relative permittivity from 1 to 3 while the cladding is assumed to be air, the maximum \(|S_3|\) values remain close to 1, and Janus chirality emission remain achievable regardless of the \(\sigma_z\) symmetry (see Figs.S5 and S6). Together with the Janus character, this distinct type of optical chirality provides a fresh perspective on the fundamental requirements of chiral light-matter interactions.

To clarify the origin of photonic eigenstates exhibiting Janus chirality, we employ the TCMT framework explained in Supplemental Information. It details a peculiar rule relating the \(\sigma_z\) mirror symmetry and the eigenstate far-field polarization: it has to be linearly polarized as long as the far fields of all contributing states are orthogonal. As the states possess certain parity (are odd and even with respect to the $z$-axis), their far fields remain orthogonal until a pair of states of the same parity happen to interfere in a certain frequency range. Then their specific coupling through the radiative damping induces hybridization and the far field can be elliptically and, eventually, even circularly polarized. 

Of particular relevance for the CWMs with Janus chirality is a combination of two even modes with substantially different decay rates: a dark mode 1 (GMR described above) and a bright electric dipole (ED) mode 2 (see Figs.S7-S10). For considerably different decay rates, $\gamma_1\ll\gamma_1$, the modes remain in the regime of weak coupling and the observable eigenstates underpinning optical resonances possess eigenfrequencies shifted by the interaction as:
\begin{equation}
    \Omega_{\pm} \approx \Omega_{1,2} \mp \frac{\gamma_{12}^2}{\Omega_1 - \Omega_2},
\end{equation}
where \(\Omega_{1,2}\) are the complex eigenfrequencies of the coupled modes, 
and the constant of their coupling $\gamma_{12}=\sqrt{\gamma_1\gamma_2}\cos\Theta$ is determined by the mode decay rates $\gamma_{1,2}$ and the angle $\Theta$ between the directions of their linearly polarized far fields.  

\begin{figure}
    \centering
    \includegraphics[width=0.75\linewidth]{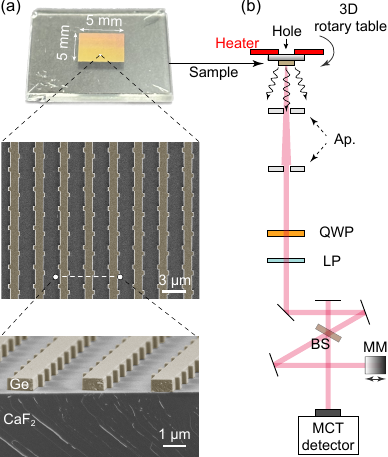}
    \caption{(a) Optical microscopy and top-view/cross-sectional SEM images of the fabricated thermal meta-emitter ($\delta$ = 735 $nm$). (b) Schematic of the experimental setup used for ThE characterization of the structure. Ap., Aperture; QWP, Quarter-wave plate; LP, Linear polarizer; BS, Beam splitter; MM, Moving mirror.}
    \label{fig3}
\end{figure}

\begin{figure*}
    \centering
    \includegraphics[width=0.8\textwidth]{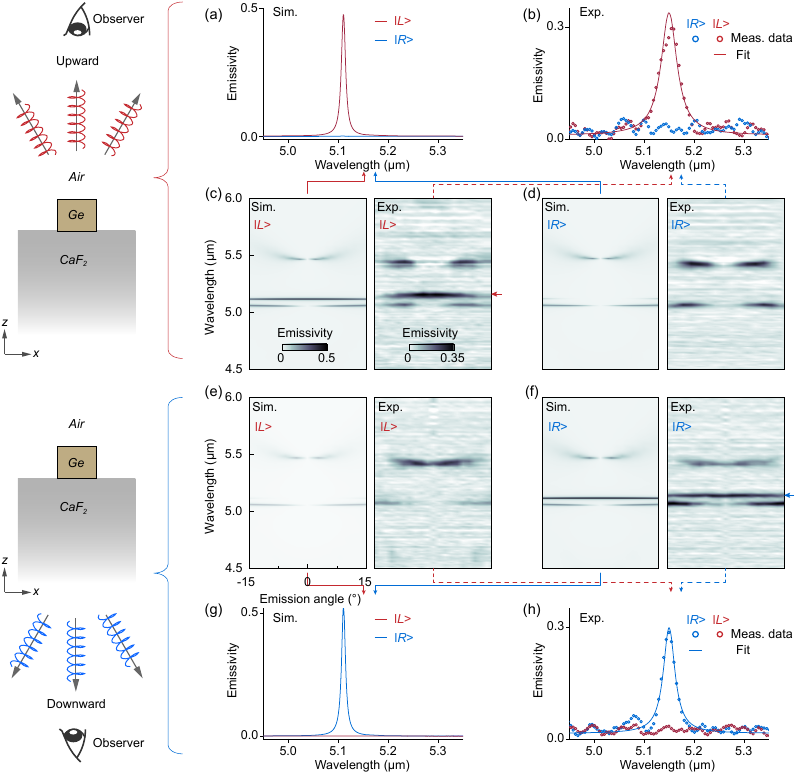}
    \caption{(a,b) Numerically simulated and experimentally measured spectra of thermal radiation in the normal direction for upward emission from the structure in Fig.~\ref{fig3}, with red and blue lines denoting LCP and RCP, respectively. (c, d) Simulated and measured angular-resolved spectra of upward emission with the opposite helicities, confirming high chirality contrast in the photonic flatband. (e, f) Simulated and measured angular-resolved spectra of downward emission with the opposite helicities, showing helicities reversed relative to upward emission and thus demonstrating Janus spin-coherent tghermal emission. (g, h) Emission spectra in the normal direction extracted from (e) and (f).}
    \label{fig4}
\end{figure*}

The coupling weakly affects the properties of the bright ED mode which becomes a low-Q state ``$-$'' remaining close to linearly polarized. The dark mode is affected much more significantly and it is transformed into a high-Q state ``$+$'' and its far field can become circularly polarized when
\begin{eqnarray}
    \sin^2\Theta =  {\gamma_1}/{\gamma_2}, \\
    \omega_2 - \omega_1  =  \sqrt{\gamma_1 \gamma_2} \cos\Theta,
\end{eqnarray}
i.e., when the polarization directions of the two modes are almost collinear and their eigenfrequencies differ accordingly. 
When these conditions are fulfilled, the high-Q state ``$+$'' becomes even narrower with the radiative decay rate $ \gamma_+=2{\gamma_1^2}/{\gamma_2}$.
In transmission, it is observed by a sharp dip of the cross-polarized transmittance: 
\begin{equation}\label{eq:TRL}
    T_{RL} = \left|\frac{1}{2}+\frac{\gamma_+}{i(\omega - \omega_+) - (\gamma_++\gamma_d)} \right|^2,
\end{equation}
only when a dissipative loss rate $\gamma_d$ is also present. At the resonance $\omega =\omega_+$, the transmittance \eqref{eq:TRL} drops to zero at the critical coupling condition $\gamma_+=\gamma_d$ which is also optimal for the thermal emission as discussed above.

Following the optimized design, we fabricated the thermal meta-emitter with $\delta$ = 735 nm using standard nanofabrication techniques (for more details, see Section 1 of the Supplemental Information). Fig.\ref{fig3}(a) presents a photograph of the sample, alongside top-view and cross-sectional false-color scanning electron microscope (SEM) images, demonstrating high structural fidelity of the fabricated structure. Fig.\ref{fig3}(b) illustrates the schematic of the customized experimental setup used for characterizing the Janus chiral thermal emission. The sample is mounted on a three-axis motorized rotating stage integrated with a heating system. Two coaxial apertures placed behind the sample are used to significantly reduce background noise, with a quarter-wave plate and a rotating linear polarizer used to resolve the circular polarization characteristics. Finally, the collected emission signal was analyzed using a Fourier-transform infrared spectrometer equipped with a mercury cadmium telluride detector.

Upon heating the sample to 200°C, we experimentally characterized the Janus chiral emission from both sides of the thermal meta-emitter, as shown in Fig.\ref{fig4}. We first measured the angle-resolved emission spectra for opposite helicities in the upward direction [Figs.\ref{fig4}(c) and (d)]. Band A exhibits exclusively LCP emission with the central wavelength remaining invariant across a broad angular range, confirming that the chiral flatband nature is in almost perfect agreement with the theoretical predictions. For upward emission in the normal direction, the calculated Q-factor and emission CD are \(Q = 545\) and \(\sim 1\), respectively, while the experimental values reach 139 and $\sim0.86$, demonstrating high temporal and spin coherence. Here, the emission CD is defined as \cite{Sun2025Circularly}: \(\text{CD} = \frac{E_{|L\rangle} - E_{|R\rangle}}{E_{|L\rangle} + E_{|R\rangle}}\), where \(E_{|L>}\) and \(E_{|R>}\) represent the emissivity for LCP and RCP outputs, respectively. Similarly, the downward emission spectra [Figs.\ref{fig4}(e-h)] reveal a $Q$-factor of 192 and a CD of \(\sim -0.8\) at normal output, confirming the handedness reversal characteristic of Janus emission. Collectively, these results provide a definitive experimental demonstration of Janus chiral thermal emission, where high-Q photonic flatbands facilitate both high temporal coherence and a rainbow-free effect. Notably, the experimental $Q$-factor is slightly lower than the calculated value, mainly due to the steep dispersion in the $y$-direction, which results in thermal emission signals from different frequencies being collected over a certain angular range. Additionally, the rough sidewalls of the sample cause extra scattering losses. The slight reduction in the CD value is mainly attributed to the unavoidable background thermal noise.

In conclusion, we have described high-$Q$ Janus-like chiral photonic eigenstates in metasurfaces composed of arrays of periodically modulated waveguides. The states are characterized by flatband dispersion and circularly polarized far fields in a broad range of the reciprocal space. This unique combination allows employing them for chiral high-coherence thermal emission characterized by a rainbow-free response across a broad rang of angles. Uniquely, such metasurfaces maintain the chiral response regardless of whether their vertical ($\sigma_z$) symmetry is intact or broken by a substrate. We have experimentally confirmed the coherent emission of thermal photons with opposite spin angular momenta into the opposite half-spaces with near-unity circular dichroism and high temporal coherence. Being supported by a coupled-mode theory, our approach offers a robust foundation for versatile chiral light-matter interactions, also paving the way for multifunctional chiral IR light sources. Beyond thermal emission, our approach could be extended to the field of valleytronics, where the distinct spin-polarized emissions from $K$ and $K'$ valleys can be spatially separated and routed into opposite directions, opening new avenues for scalable quantum information processing.

\begin{acknowledgments}
This work was supported by the National Natural Science Foundation of China (Grant No. 12274269) and the University of Hong Kong through the Hung Hing Ying Distinguished Visiting Professorship.
\end{acknowledgments}

\bibliography{Ref}

@PREAMBLE{
 "\providecommand{\noopsort}[1]{}" 
 # "\providecommand{\singleletter}[1]{#1}%" 
}

@article{Huang2023Resonant,
  title = {Resonant leaky modes in all-dielectric metasystems: Fundamentals and applications},
  author = {Huang,  Lujun and Xu,  Lei and Powell,  David A. and Padilla,  Willie J. and Miroshnichenko,  Andrey E.},
  journal = {Phys. Rep.},
  volume = {1008},
  ISSN = {0370-1573},
  pages = {1–66},
  year = {2023},
  url = {http://dx.doi.org/10.1016/j.physrep.2023.01.001},
  DOI = {10.1016/j.physrep.2023.01.001},
  publisher = {Elsevier BV},
}

@article{Chen2025Thermal,
  title = {Thermal Digital Imaging by Chirally Coded Metasurfaces},
  author = {Chen,  Tianle and Raza,  Faizan and Ni,  Chen and Zhang,  Gan and Chen,  Rui and Li,  Xinran and Dang,  Yongdi and Zhang,  Sen and Ng,  Chong‐Kuong and Hu,  Jun and Di,  Dawei and Choudhury,  Pankaj K. and Ma,  Yungui},
  journal = {Laser Photon. Rev.},
  volume = {19},
  number = {11},
  pages={2500129},
  year = {2025},
  url = {http://dx.doi.org/10.1002/lpor.202500129},
  DOI = {10.1002/lpor.202500129},
  publisher = {Wiley}
}

@article{Gorkunov2020Metasurfaces,
  title = {Metasurfaces with Maximum Chirality Empowered by Bound States in the Continuum},
  author = {Gorkunov,  Maxim V. and Antonov,  Alexander A. and Kivshar,  Yuri S.},
  journal = {Phys. Rev. Lett.},
  volume = {125},
  number = {9},
  pages={093903},
  year = {2020},
  url = {http://dx.doi.org/10.1103/physrevlett.125.093903},
  DOI = {10.1103/physrevlett.125.093903},
  publisher = {American Physical Society (APS)},
}

@article{Miyoshi2018High,
  title = {High-speed and on-chip graphene blackbody emitters for optical communications by remote heat transfer},
  author = {Miyoshi,  Yusuke and Fukazawa,  Yusuke and Amasaka,  Yuya and Reckmann,  Robin and Yokoi,  Tomoya and Ishida,  Kazuki and Kawahara,  Kenji and Ago,  Hiroki and Maki,  Hideyuki},
  journal = {Nat. Commun.},
  volume = {9},
  number = {1},
  pages = {1279},
  year = {2018},
  url = {http://dx.doi.org/10.1038/s41467-018-03695-x},
  DOI = {10.1038/s41467-018-03695-x},
  publisher = {Springer Science and Business Media LLC},
}

@article{DeZoysa2012Conversion,
  title = {Conversion of broadband to narrowband thermal emission through energy recycling},
  author = {De Zoysa,  Menaka and Asano,  Takashi and Mochizuki,  Keita and Oskooi,  Ardavan and Inoue,  Takuya and Noda,  Susumu},
  journal = {Nat. Photonics},
  volume = {6},
  number = {8},
  pages = {535–539},
  year = {2012},
  url = {http://dx.doi.org/10.1038/NPHOTON.2012.146},
  DOI = {10.1038/nphoton.2012.146},
  publisher = {Springer Science and Business Media LLC},
}

@article{Tan2020Non,
  title = {Non-dispersive infrared multi-gas sensing via nanoantenna integrated narrowband detectors},
  author = {Tan,  Xiaochao and Zhang,  Heng and Li,  Junyu and Wan,  Haowei and Guo,  Qiushi and Zhu,  Houbin and Liu,  Huan and Yi,  Fei},
  journal = {Nat. Commun.},
  volume = {11},
  number = {1},  
  pages = {5245},
  year = {2020},
  url = {http://dx.doi.org/10.1038/s41467-020-19085-1},
  DOI = {10.1038/s41467-020-19085-1},
  publisher = {Springer Science and Business Media LLC},
}

@article{Yariv1999,
  title = {Coupled-resonator optical waveguide:a proposal and analysis},
  author = {Yariv,  Amnon and Xu,  Yong and Lee,  Reginald K. and Scherer,  Axel},
  journal = {Opt. Lett.},
  volume = {24},
  number = {11},
  pages = {711},
  year = {1999},
  url = {http://dx.doi.org/10.1364/OL.24.000711},
  DOI = {10.1364/ol.24.000711},
  publisher = {Optica Publishing Group},
}

@article{Chen2023Directional,
  title = {Directional terahertz holography with thermally active Janus metasurface},
  author = {Chen,  Benwen and Yang,  Shengxin and Chen,  Jian and Wu,  Jingbo and Chen,  Ke and Li,  Weili and Tan,  Yihui and Wang,  Zhaosong and Qiu,  Hongsong and Fan,  Kebin and Zhang,  Caihong and Wang,  Huabing and Feng,  Yijun and He,  Yunbin and Jin,  Biaobing and Wu,  Xinglong and Chen,  Jian and Wu,  Peiheng},
  journal = {Light Sci. Appl.},
  volume = {12},
  number = {1},
  pages = {136},
  year = {2023},
  url = {http://dx.doi.org/10.1038/s41377-023-01177-4},
  DOI = {10.1038/s41377-023-01177-4},
  publisher = {Springer Science and Business Media LLC},
}

@article{Zhao2024Spin,
  title = {Spin-Orbit-Locking Chiral Bound States in the Continuum},
  author = {Zhao,  Xingqi and Wang,  Jiajun and Liu,  Wenzhe and Che,  Zhiyuan and Wang,  Xinhao and Chan,  C. T. and Shi,  Lei and Zi,  Jian},
  journal = {Phys. Rev. Lett.},
  volume = {133},
  number = {3},
  pages = {036201},
  year = {2024},
  url = {http://dx.doi.org/10.1103/PhysRevLett.133.036201},
  DOI = {10.1103/physrevlett.133.036201},
  publisher = {American Physical Society (APS)},
}

@article{Lv2024Robust,
  title = {Robust generation of intrinsic C points with magneto-optical bound states in the continuum},
  author = {Lv,  Wenjing and Qin,  Haoye and Su,  Zengping and Zhang,  Chengzhi and Huang,  Jiongpeng and Shi,  Yuzhi and Li,  Bo and Genevet,  Patrice and Song,  Qinghua},
  journal = {Sci. Adv.},
  volume = {10},
  number = {46},
  pages = {eads0157},
  year = {2024},
  url = {http://dx.doi.org/10.1126/sciadv.ads0157},
  DOI = {10.1126/sciadv.ads0157},
  publisher = {American Association for the Advancement of Science (AAAS)},
}

@article{Wang2024Optical,
  title = {Optical bound states in the continuum in periodic structures: mechanisms,  effects,  and applications},
  author = {Wang,  Jiajun and Li,  Peishen and Zhao,  Xingqi and Qian,  Zhiyuan and Wang,  Xinhao and Wang,  Feifan and Zhou,  Xinyi and Han,  Dezhuan and Peng,  Chao and Shi,  Lei and Zi,  Jian},
  journal = {Photonics Insights},
  volume = {3},
  number = {1},
  pages = {R01},
  year = {2024},
  url = {http://dx.doi.org/10.3788/PI.2024.R01},
  DOI = {10.3788/pi.2024.r01},
  publisher = {Shanghai Institute of Optics and Fine Mechanics},
}

@article{Liu2019Circularly,
  title = {Circularly Polarized States Spawning from Bound States in the Continuum},
  author = {Liu,  Wenzhe and Wang,  Bo and Zhang,  Yiwen and Wang,  Jiajun and Zhao,  Maoxiong and Guan,  Fang and Liu,  Xiaohan and Shi,  Lei and Zi,  Jian},
  journal = {Phys. Rev. Lett.},
  volume = {123},
  number = {11},
  pages = {116104},
  year = {2019},
  url = {http://dx.doi.org/10.1103/PhysRevLett.123.116104},
  DOI = {10.1103/physrevlett.123.116104},
  publisher = {American Physical Society (APS)},
}

@article{Picardi2018Janus,
  title = {Janus and Huygens Dipoles: Near-Field Directionality Beyond Spin-Momentum Locking},
  author = {Picardi,  Michela F. and Zayats,  Anatoly V. and Rodríguez-Fortuño,  Francisco J.},
  journal = {Phys. Rev. Lett.},
  volume = {120},
  number = {11},
  pages = {117402},
  year = {2018},
  url = {http://dx.doi.org/10.1103/PhysRevLett.120.117402},
  DOI = {10.1103/physrevlett.120.117402},
  publisher = {American Physical Society (APS)},
}

@article{Kang2025Janus,
  title = {Janus Bound States in the Continuum with Asymmetric Topological Charges},
  author = {Kang,  Meng and Xiao,  Meng and Chan,  C.T.},
  journal = {Phys. Rev. Lett.},
  volume = {134},
  number = {1},
  pages = {013805},
  year = {2025},
  url = {http://dx.doi.org/10.1103/PhysRevLett.134.013805},
  DOI = {10.1103/physrevlett.134.013805},
  publisher = {American Physical Society (APS)},
}

@article{Zuo2025Janus,
  title = {Janus bound states in the continuum in structurally symmetric photonic crystals},
  author = {Zuo,  Hongzhi and Xia,  Shengxuan and Meng,  Haiyu},
  journal = {Phys. Rev. B.},
  volume = {112},
  number = {16},
  pages = {165414},
  year = {2025},
  url = {http://dx.doi.org/10.1103/s2zt-bbvn},
  DOI = {10.1103/s2zt-bbvn},
  publisher = {American Physical Society (APS)},
}

@article{Lin2025Chiral,
  title = {Chiral Metasurfaces with Multidimensional Tunability for Optical Chiral States},
  author = {Lin,  Dayang and Ghaffari,  Abbas and Gu,  Qing},
  journal = {Adv. Opt. Mater.},
  volume = {13},
  number = {36},
  pages = {e01859},
  year = {2025},
  url = {http://dx.doi.org/10.1002/adom.202501859},
  DOI = {10.1002/adom.202501859},
  publisher = {Wiley},
}

@article{Ma2026Janus,
  title = {Janus bound states in the continuum with complex asymmetric topological charges},
  author = {Ma,  Ke and Ma,  Yumeng and Deng,  Fangze and Han,  Zhihua and Liu,  Meng and Zhang,  Jinjuan and Zhang,  Yuping and Zhang,  Huiyun},
  journal = {J. Appl. Phys.},
  volume = {139},
  number = {1},
  pages = {013104},
  year = {2026},
  url = {http://dx.doi.org/10.1063/5.0310977},
  DOI = {10.1063/5.0310977},
  publisher = {AIP Publishing},
}

@article{Yin2025Janus,
  title = {Janus chiral bound states in the continuum with maximum circular dichroism in photonic crystal heterostructures},
  author = {Yin,  Zhibin and Cao,  Zhaolong},
  journal = {Opt. Lett.},
  volume = {50},
  number = {18},
  pages = {5710},
  year = {2025},
  url = {http://dx.doi.org/10.1364/OL.568029},
  DOI = {10.1364/ol.568029},
  publisher = {Optica Publishing Group},
}

@article{Song2025Parity,
  title = {Parity-time-symmetry-induced Janus bound states in the continuum and rings of lasing threshold modes},
  author = {Song,  Qianju and Yi,  Zao and Xiang,  Hong and Han,  Dezhuan},
  journal = {Phys. Rev. A},
  volume = {112},
  number = {6},
  pages = {063517},
  year = {2025},
  url = {http://dx.doi.org/10.1103/37yr-5pfx},
  DOI = {10.1103/37yr-5pfx},
  publisher = {American Physical Society (APS)},
}

@article{Wang2025High,
  title = {High‐Q Chiral Janus Metasurfaces Based on Multipolar Resonances},
  author = {Wang,  Chaoyang and Wu,  Jiaju and Tian,  Jingyi},
  journal = {Laser Photon. Rev.},
  volume = {19},
  number = {21},
  pages = {e00205},
  year = {2025},
  url = {http://dx.doi.org/10.1002/lpor.202500205},
  DOI = {10.1002/lpor.202500205},
  publisher = {Wiley},
}

@article{Ji2026Janus,
  title = {Janus bound states in the continuum and robust unidirectional guided resonances induced by shear},
  author = {Ji,  Chang-Yin and Nguyen,  Hai Son and Hu,  Guangwei},
  journal = {Newton},
  volume = {2},
  number = {1},
  pages = {100318},
  year = {2026},
  url = {http://dx.doi.org/10.1016/j.newton.2025.100318},
  DOI = {10.1016/j.newton.2025.100318},
  publisher = {Elsevier BV},
}

@article{Cotrufo2023nonreciprocal,
  title = {Passive bias-free non-reciprocal metasurfaces based on thermally nonlinear quasi-bound states in the continuum},
  author = {Cotrufo,  Michele and Cordaro,  Andrea and Sounas,  Dimitrios L. and Polman,  Albert and Alù,  Andrea},
  journal = {Nat. Photonics},
  volume = {18},
  number = {1},
  pages = {81–90},
  year = {2023},
  url = {http://dx.doi.org/10.1038/s41566-023-01333-7},
  DOI = {10.1038/s41566-023-01333-7},
  publisher = {Springer Science and Business Media LLC},
}

@article{Zhu2021Janus,
  title = {Janus acoustic metascreen with nonreciprocal and reconfigurable phase modulations},
  author = {Zhu,  Yifan and Cao,  Liyun and Merkel,  Aurélien and Fan,  Shi-Wang and Vincent,  Brice and Assouar,  Badreddine},
  journal = {Nat. Commun.},
  volume = {12},
  number = {1},
  pages = {7089},
  year = {2021},
  url = {http://dx.doi.org/10.1038/s41467-021-27403-4},
  DOI = {10.1038/s41467-021-27403-4},
  publisher = {Springer Science and Business Media LLC},
}

@article{Wang2025Terahertz,
  title = {Terahertz Nonvolatile Vectorial Holography With Random Repetitive Encryption and Nonreciprocal Janus Imaging Based on Magnetic Heterogeneous Integrated Metasurface},
  author = {Wang,  Hao and Fan,  Fei and Li,  Pengxuan and Xue,  Qiang and Tan,  Zhiyu and Zhao,  Dan and Zhao,  Huijun and Yang,  Qinghui and Wen,  Qiye and Chang,  Shengjiang},
  journal = {Laser Photon. Rev.},
  volume = {19},
  number = {14},
  pages = {2500375},
  year = {2025},
  url = {http://dx.doi.org/10.1002/lpor.202500375},
  DOI = {10.1002/lpor.202500375},
  publisher = {Wiley},
}

@article{Chen2019Directional,
  title = {Directional Janus Metasurface},
  author = {Chen,  Ke and Ding,  Guowen and Hu,  Guangwei and Jin,  Zhongwei and Zhao,  Junming and Feng,  Yijun and Jiang,  Tian and Alù,  Andrea and Qiu,  Cheng‐Wei},
  journal = {Adv. Mater.},
  volume = {32},
  number = {2},
  pages = {1906352},
  year = {2019},
  url = {http://dx.doi.org/10.1002/adma.201906352},
  DOI = {10.1002/adma.201906352},
  publisher = {Wiley},
}

@article{Sun2025Circularly,
  title = {Circularly polarized thermal emission driven by chiral flatbands in monoclinic metasurfaces},
  author = {Sun,  Kaili and Yang,  Bingxiong and Cai,  Yangjian and Kivshar,  Yuri and Han,  Zhanghua},
  journal = {Sci. Adv.},
  volume = {11},
  number = {31},
  pages = {adw0986},
  year = {2025},
  url = {http://dx.doi.org/10.1126/sciadv.adw0986},
  DOI = {10.1126/sciadv.adw0986},
  publisher = {American Association for the Advancement of Science (AAAS)},
}

@article{Sun2024Ultra,
  title = {Ultra-narrowband and rainbow-free mid-infrared thermal emitters enabled by a flat band design in distorted photonic lattices},
  author = {Sun,  Kaili and Cai,  Yangjian and Huang,  Lujun and Han,  Zhanghua},
  journal = {Nat. Commun.},
  volume = {15},
  number = {1},
  pages = {4019},
  year = {2024},
  url = {http://dx.doi.org/10.1038/s41467-024-48499-4},
  DOI = {10.1038/s41467-024-48499-4},
  publisher = {Springer Science and Business Media LLC},
}

@article{Sun2025Full,
  title = {Full polarization and high coherence control of thermal emissions via saddle-band dispersion engineering},
  author = {Sun,  Kaili and Wang,  Guangdong and Li,  Wenyu and Wang,  Yinghan and Cai,  Yangjian and Huang,  Lujun and Alù,  Andrea and Han,  Zhanghua},
  journal = {Nat. Commun.},
  volume = {16},
  number = {1},
  pages = {8393},
  year = {2025},
  url = {http://dx.doi.org/10.1038/s41467-025-63334-0},
  DOI = {10.1038/s41467-025-63334-0},
  publisher = {Springer Science and Business Media LLC},
}

@article{Wang2023Observation,
  title = {Observation of nonvanishing optical helicity in thermal radiation from symmetry-broken metasurfaces},
  author = {Wang,  Xueji and Sentz,  Tyler and Bharadwaj,  Sathwik and Ray,  Subir Kumar and Wang,  Yifan and Jiao,  Dan and Qi,  Limei and Jacob,  Zubin},
  journal = {Sci. Adv.},
  volume = {9},
  number = {4},
  pages = {eade4203},
  year = {2023},
  url = {http://dx.doi.org/10.1126/sciadv.ade4203},
  DOI = {10.1126/sciadv.ade4203},
  publisher = {American Association for the Advancement of Science (AAAS)},
}

@article{Greffet2002Coherent,
  title = {Coherent emission of light by thermal sources},
  author = {Greffet,  Jean-Jacques and Carminati,  Rémi and Joulain,  Karl and Mulet,  Jean-Philippe and Mainguy,  Stéphane and Chen,  Yong},
  journal = {Nature},
  volume = {416},
  number = {6876},
  pages = {61–64},
  year = {2002},
  url = {http://dx.doi.org/10.1038/416061a},
  DOI = {10.1038/416061a},
  publisher = {Springer Science and Business Media LLC},
}

@article{Inoue2014Realization,
  title = {Realization of dynamic thermal emission control},
  author = {Inoue,  Takuya and Zoysa,  Menaka De and Asano,  Takashi and Noda,  Susumu},
  journal = {Nat. Mater.},
  volume = {13},
  number = {10},
  pages = {928–931},
  year = {2014},
  url = {http://dx.doi.org/10.1038/NMAT4043},
  DOI = {10.1038/nmat4043},
  publisher = {Springer Science and Business Media LLC},
}

@article{Nolen2024Local,
  title = {Local control of polarization and geometric phase in thermal metasurfaces},
  author = {Nolen,  J. Ryan and Overvig,  Adam C. and Cotrufo,  Michele and Alù,  Andrea},
  journal = {Nat. Nanotechnol.},
  volume = {19},
  number = {11},
  pages = {1627–1634},
  year = {2024},
  url = {http://dx.doi.org/10.1038/s41565-024-01763-6},
  DOI = {10.1038/s41565-024-01763-6},
  publisher = {Springer Science and Business Media LLC},
}

@article{Overvig2021Thermal,
  title = {Thermal Metasurfaces: Complete Emission Control by Combining Local and Nonlocal Light-Matter Interactions},
  author = {Overvig,  Adam C. and Mann,  Sander A. and Alù,  Andrea},
  journal = {Phys. Rev. X.},
  volume = {11},
  number = {2},
  pages = {021050},
  year = {2021},
  url = {http://dx.doi.org/10.1103/PhysRevX.11.021050},
  DOI = {10.1103/physrevx.11.021050},
  publisher = {American Physical Society (APS)},
}

@article{Zhang2022Chiral,
  title = {Chiral emission from resonant metasurfaces},
  author = {Zhang,  Xudong and Liu,  Yilin and Han,  Jiecai and Kivshar,  Yuri and Song,  Qinghai},
  journal = {Science},
  volume = {377},
  number = {6611},
  pages = {1215–1218},
  year = {2022},
  url = {http://dx.doi.org/10.1126/science.abq7870},
  DOI = {10.1126/science.abq7870},
  publisher = {American Association for the Advancement of Science (AAAS)},
}

@article{Sun2025Flatband,
  title = {Flatband high-Q metasurfaces inspired by coupled-resonator optical waveguides},
  author = {Sun,  Kaili and Cai,  Yangjian and Kivshar,  Yuri and Han,  Zhanghua},
  journal = {Adv. Photonics},
  volume = {7},
  number = {05},
  pages = {056008},
  year = {2025},
  url = {http://dx.doi.org/10.1117/1.AP.7.5.056008},
  DOI = {10.1117/1.ap.7.5.056008},
  publisher = {SPIE-Intl Soc Optical Eng},
}

@article{Brongersma2025The,
  title = {The second optical metasurface revolution: moving from science to technology},
  author = {Brongersma,  Mark L. and Pala,  Ragip A. and Altug,  Hatice and Capasso,  Federico and Chen,  Wei Ting and Majumdar,  Arka and Atwater,  Harry A.},
  journal = {Nat. Rev. Electr. Eng.},
  volume = {2},
  number = {2},
  pages = {125–143},
  year = {2025},
  url = {http://dx.doi.org/10.1038/s44287-024-00136-4},
  DOI = {10.1038/s44287-024-00136-4},
  publisher = {Springer Science and Business Media LLC},
}

@article{Shi2022Planar,
  title = {Planar chiral metasurfaces with maximal and tunable chiroptical response driven by bound states in the continuum},
  author = {Shi,  Tan and Deng,  Zi-Lan and Geng,  Guangzhou and Zeng,  Xianzhi and Zeng,  Yixuan and Hu,  Guangwei and Overvig,  Adam and Li,  Junjie and Qiu,  Cheng-Wei and Alù,  Andrea and Kivshar,  Yuri S. and Li,  Xiangping},
  journal = {Nat. Commun.},
  volume = {13},
  number = {1},
  pages = {4111},
  year = {2022},
  url = {http://dx.doi.org/10.1038/s41467-022-31877-1},
  DOI = {10.1038/s41467-022-31877-1},
  publisher = {Springer Science and Business Media LLC},
}

@article{Chen2023Observation,
  title = {Observation of intrinsic chiral bound states in the continuum},
  author = {Chen,  Yang and Deng,  Huachun and Sha,  Xinbo and Chen,  Weijin and Wang,  Ruize and Chen,  Yu-Hang and Wu,  Dong and Chu,  Jiaru and Kivshar,  Yuri S. and Xiao,  Shumin and Qiu,  Cheng-Wei},
  journal = {Nature},
  volume = {613},
  number = {7944},
  pages = {474–478},
  year = {2023},
  url = {http://dx.doi.org/10.1038/s41586-022-05467-6},
  DOI = {10.1038/s41586-022-05467-6},
  publisher = {Springer Science and Business Media LLC},
}

@article{Sun2023Infinite,
  title = {Infinite-Q guided modes radiate in the continuum},
  author = {Sun,  Kaili and Wei,  Heng and Chen,  Weijin and Chen,  Yang and Cai,  Yangjian and Qiu,  Cheng-Wei and Han,  Zhanghua},
  journal = {Phys. Rev. B.},
  volume = {107},
  number = {11},
  pages = {115415},
  year = {2023},
  url = {http://dx.doi.org/10.1103/PhysRevB.107.115415},
  DOI = {10.1103/physrevb.107.115415},
  publisher = {American Physical Society (APS)},
}

@article{Sun2024High,
  title = {High-Q resonances in periodic photonic structures},
  author = {Sun,  Kaili and Wang,  Wei and Han,  Zhanghua},
  journal = {Phys. Rev. B.},
  volume = {109},
  number = {8},
  pages = {085426},
  year = {2024},
  url = {http://dx.doi.org/10.1103/PhysRevB.109.085426},
  DOI = {10.1103/physrevb.109.085426},
  publisher = {American Physical Society (APS)},
}

@article{Kang2021Merging,
  title = {Merging Bound States in the Continuum at Off-High Symmetry Points},
  author = {Kang,  Meng and Zhang,  Shunping and Xiao,  Meng and Xu,  Hongxing},
  journal = {Phys. Rev. Lett.},
  volume = {126},
  number = {11},
  pages = {085426},
  year = {2021},
  url = {http://dx.doi.org/10.1103/PhysRevLett.126.117402},
  DOI = {10.1103/physrevlett.126.117402}, 
  publisher = {American Physical Society (APS)},
}

@article{Jin2019Topologically,
  title = {Topologically enabled ultrahigh-Q guided resonances robust to out-of-plane scattering},
  author = {Jin,  Jicheng and Yin,  Xuefan and Ni,  Liangfu and Soljačić,  Marin and Zhen,  Bo and Peng,  Chao},
  journal = {Nature},
  volume = {574},
  number = {7779},
  pages = {501–504},
  year = {2019},
  url = {http://dx.doi.org/10.1038/s41586-019-1664-7},
  DOI = {10.1038/s41586-019-1664-7},
  publisher = {Springer Science and Business Media LLC},
}

@article{Ye2020Singular,
  title = {Singular Points of Polarizations in the Momentum Space of Photonic Crystal Slabs},
  author = {Ye,  Weimin and Gao,  Yang and Liu,  Jianlong},
  journal = {Phys. Rev. Lett.},
  volume = {124},
  number = {15},
  pages = {153904},
  year = {2020},
  url = {http://dx.doi.org/10.1103/PhysRevLett.124.153904},
  DOI = {10.1103/physrevlett.124.153904},
  publisher = {American Physical Society (APS)},
}

@article{Zhen2014Topological,
  title = {Topological Nature of Optical Bound States in the Continuum},
  author = {Zhen,  Bo and Hsu,  Chia Wei and Lu,  Ling and Stone,  A. Douglas and Soljačić,  Marin},
  journal = {Phys. Rev. Lett.},
  volume = {113},
  number = {25},
  pages = {257401},
  year = {2014},
  url = {http://dx.doi.org/10.1103/PhysRevLett.113.257401},
  DOI = {10.1103/physrevlett.113.257401},
  publisher = {American Physical Society (APS)},
}

@article{Kuehner2023,
 title     = {Unlocking the out-of-plane dimension for photonic bound states in the continuum to achieve maximum optical chirality},
  author    = {Kühner, Lucca and Wendisch, Fedja J. and Antonov, Alexander A. and Bürger, Johannes and Hüttenhofer, Ludwig and de S. Menezes, Leonardo and Maier, Stefan A. and Gorkunov, Maxim V. and Kivshar, Yuri and Tittl, Andreas},
  journal   = {Light: Science \& Applications},
  volume = {12},
  number = {1},
  pages = {250},
  year = {2023},
  url = {http://dx.doi.org/10.1038/s41377-023-01295-z},
  DOI = {10.1038/s41377-023-01295-z},
  publisher = {Springer Science and Business Media LLC},
  
}

@misc{Heimig2024,
Author = {Connor Heimig and Alexander A. Antonov and Dmytro Gryb and Thomas Possmayer and Thomas Weber and Michael Hirler and Jonas Biechteler and Luca Sortino and Leonardo de S. Menezes and Stefan A. Maier and Maxim V. Gorkunov and Yuri Kivshar and Andreas Tittl},
Title = {Chiral Nonlinear Polaritonics with van der Waals Metasurfaces},
Year = {2024},
Eprint = {arXiv:2410.18760},
}

@Book{Haus1984,
  title       = {Waves and fields in optoelectronics},
  publisher   = {Englewood Cliffs, NJ: Prentice-Hall},
  year        = {1984},
  author      = {Haus, Hermann A.},
  call-number = {TA1750 .H38 1984},
  comment     = {&quot;The introductory graduate course &#39;Optics and optical electronics,&#39; the notes of which developed into this book, was taught jointly by Professor S. Ezekiel and the author&quot;-- P. xii.;Includes bibliographical references and indexes.},
  refid       = {999550742402121},
  url         = {https://search.library.wisc.edu/catalog/999550742402121},
}

\end{document}